\UseRawInputEncoding
\documentclass[%
 reprint,
amsmath,amssymb,
aps,
prb,
]{revtex4-2}

\usepackage{graphicx}
\usepackage{dcolumn}
\usepackage{bm}

\makeatletter

\newcommand{\Rmnum}[1]{\expandafter\@slowromancap\romannumeral #1@}
\makeatother
\usepackage{float}
\usepackage{amsthm}
\usepackage{mathrsfs}
\usepackage{amssymb}
\usepackage{epsfig}
\usepackage{amsmath}
\usepackage{epsfig}
\usepackage{booktabs}
\usepackage{multirow}
\usepackage[colorlinks,urlcolor=blue]{hyperref}
\usepackage{makecell}
\usepackage[utf8]{inputenc}
\usepackage[section]{placeins}

\begin{document}

\preprint{APS/123-QED}
\title{Manipulation of magnetic topological textures via perpendicular strain and polarization in van der Waals magnetoelectric heterostructure}
\author{Zhong Shen}
\author{Shuai Dong}
\email[]{sdong@seu.edu.cn}
\author{Xiaoyan Yao}
\email[]{yaoxiaoyan@seu.edu.cn}

\affiliation{Key Laboratory of Quantum Materials and Devices of Ministry of Education, School of Physics, Southeast University, Nanjing 211189, China}



\date{\today}

\begin{abstract}
Multi-functional manipulation of magnetic topological textures such as skyrmions and bimerons in energy-efficient ways is of great importance for spintronic applications, but still being a big challenge. Here, by first-principles calculations and atomistic simulations, the creation and annihilation of skyrmions/bimerons, as key operations for the reading and writing of information in spintronic devices, are achieved in van der Waals magnetoelectric CrISe/In$_2$Se$_3$ heterostructure via perpendicular strain or electric field without external magnetic field. Besides, the bimeron-skyrmion conversion, size modulation and the reversible magnetization switching from in-plane to out-of-plane could also be realized in magnetic-field-free ways. Moreover, the topological charge and morphology can be precisely controlled by a small magnetic field. The strong Dzyaloshinskii-Moriya interaction and tunable magnetic anisotropy energy in a wide window are found to play vital roles in such energy-efficient multi-functional manipulation, and the underlying physical mechanisms are elucidated.
Our work predicts the CrISe/In$_2$Se$_3$ heterostructure being an ideal platform to address this challenge in spintronic applications, and theoretically guides the low-dissipation multi-functional manipulation of magnetic topological textures.
\end{abstract}

\maketitle

\textit{Introduction}.
Magnetic skyrmions are topologically protected spin textures with high stability against defects and disorder~\cite{RN5460,RN5461,RN5321,RN5322}. Since the first experimental observation in 2009~\cite{RN5054}, skyrmions have been investigated intensively both experimentally~\cite{RN3526,RN3723,RN3103,RN3686,RN2979,RN3177} and theoretically~\cite{RN2987,RN5284,RN109,RN3688,RN3713,RN5251,RN5293,RN5082,RN5231} due to their promising applications in next-generation nonvolatile spintronic devices with high storage density and low energy dissipation. As the counterpart of skyrmions in easy-plane magnets, bimerons~\cite{RN5300,RN3166,RN5385} can also be used as the information carriers for memory devices. Two key parameters for the stability of skyrmions and bimerons are the Dzyaloshinskii-Moriya interaction (DMI)~\cite{RN130,RN131} and the magnetic anisotropy, both of which originate from the spin orbit coupling.

For the practical applications, it is extremely important to manipulate the size and morphology of the topological textures, and to control their generation and annihilation in a convenient and low-dissipation way. Compared to the widely used traditional methods with energy dissipation, such as spin-polarized current~\cite{PhysRevLett.93.127204,SLONCZEWSKI1996L1,PhysRevB.54.9353}, thermal excitation~\cite{PhysRevLett.111.067203,RN3121}, and external magnetic field pulse~\cite{PhysRevLett.98.117201,PhysRevB.79.224429}, the strain~\cite{RN5462,RN5463} and electric field~\cite{RN5244,RN3166,RN5323,RN5324} are more energy-efficient ways to tune magnetism. Recent works have demonstrated the creation and
annihilation of bimerons~\cite{RN5300} as well as the conversion from loops of vortices and antivortices to skyrmions~\cite{RN5244}, both of which are induced by the change of magnetic anisotropy energy (MAE). However, in above works, the MAE jumps between specific values, rather than changes near-continuously in a wide range, which limits the further regulations of skyrmions and bimerons. Thus, the multi-functional manipulation of magnetic topological textures remains to be investigated and potential magnetoelectric systems are yet to be discovered.

Here, by first-principles calculations and atomistic simulations, we theoretically demonstrate the multi-functional manipulation of magnetic skyrmions and bimerons in van der Waals magnetoelectric heterostructure CrISe/In$_2$Se$_3$. The generation and annihilation, size modulation, and the bimeron-skyrmion conversion could all be realized by applying perpendicular strain or electric field without any external magnetic field. The MAE tunable in a wide range from $-$3.5 meV to 1.6 meV, namely the anisotropy switchable from easy-plane (in-plane) to easy-axis (out-of-plane), is found to play a crucial role in this multi-functional manipulation of skyrmions and bimerons.

\textit{Methods}. 
The first-principles calculations are carried out by using the Vienna $ab$ $initio$ simulation package (VASP)~\cite{RN2004} based on the density functional theory. The phonon band structures are calculated by using the PHONOPY code~\cite{Togo2015,Togo2008} with a 3$\times$3$\times$1 supercell to confirm the dynamic stability of CrISe. The visualization of crystal structures is realized by VESTA~\cite{VESTA2011}. 
To explore the topological magnetic textures of CrISe/In$_2$Se$_3$, we perform atomistic simulations based on the Heisenberg model and Landau-Lifshitz-Gilbert (LLG) equation~\cite{RN1984,RN1985} as implemented in the SPIRIT package~\cite{RN1055}. See Supplemental Material~\cite{SupplementalMaterials} for more details.

The magnetic properties of CrISe/In$_2$Se$_3$ are investigated based on the following Hamiltonian:
\begin{eqnarray}
	H =&&J_{1}\sum\limits_{\left \langle i,j \right \rangle}\vec{S_{i}}\cdot\vec{S_{j}} + J_{2}\sum\limits_{\left \langle \left \langle i,k \right \rangle \right \rangle}\vec{S_{i}}\cdot\vec{S_{k}} + J_{3}\sum\limits_{\left \langle \left \langle \left \langle i,l \right \rangle \right \rangle \right \rangle}\vec{S_{i}}\cdot\vec{S_{l}}\nonumber \\
	&&+ \sum\limits_{\left \langle i,j \right \rangle}\vec{d_{ij}}\cdot(\vec{S_{i}}\times\vec{S_{j}}) + K\sum_{i}(S_{i}^{z})^2 \nonumber \\
	&&+ \mu_{Cr}\sum\limits_{i}\vec{B}\cdot\vec{S_i}\label{2}.
\end{eqnarray}
Here, the first three terms describe the isotropic Heisenberg exchanges with $J_{1}$, $J_{2}$, $J_{3}$ being the exchange coefficients between the nearest-neighbor (NN), second NN and third NN Cr atoms. $\vec{S_{i}}$ is
a unit vector representing the orientation of the spin at the $i$th Cr atom. The DMI, magnetic anisotropy and external magnetic field are characterized by the parameters $\vec{d_{ij}}$, $K$ and $\vec{B}$ respectively. $\mu_{Cr}$ is the magnetic moment of Cr atoms. All the magnetic parameters are extracted from the energy differences of distinct magnetic configurations (See Supplemental Material~\cite{SupplementalMaterials} for more details).

\begin{figure}[!htbp] 
	\begin{center}
		\includegraphics[width=8.5cm]{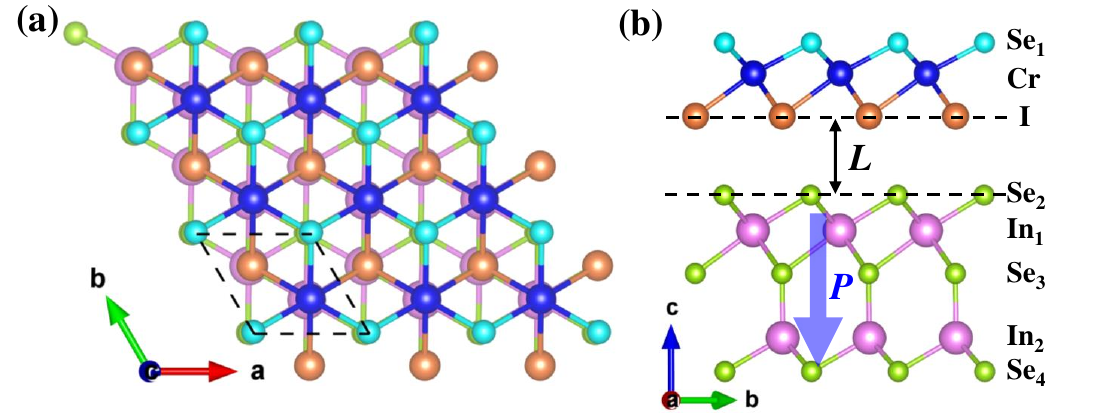}\
		\caption{\label{fig1}
		(a) Top and (b) side views of CrISe/In$_2$Se$_3$ heterostructure. The black dotted lines in (a) show the primitive cell, and the interlayer distance $L$ is defined as the distance between the atomic planes of I and Se$_2$ as shown in (b).}
\end{center}
\end{figure}
\textit{Results and discussion}. 
The In$_2$Se$_3$ monolayer is a room-temperature ferroelectric material with  reversible spontaneous electric polarization~\cite{RN5327,RN5326}. The CrISe monolayer is predicted to be a room-temperature ferromagnetic semiconductor with intrinsic broken inversion symmetry due to the Janus structure~\cite{RN5231}. The dynamical stability of CrISe monolayer is verified by the phonon spectrum calculation in Fig. S3(c)~\cite{SupplementalMaterials}.  Considering that vdW monolayers are easy to stack together, many two-dimensional (2D) vdW heterostructures have been experimentally synthesized such as MoS$_2$/In$_2$Se$_3$~\cite{RN5338}, WSe$_2$/In$_2$Se$_3$~\cite{RN5339} and FePS$_3$/Fe$_3$GeTe$_2$~\cite{RN879}. Here, we construct the CrISe/In$_2$Se$_3$ heterostructure with the ferroelectric polarization ($P$) of In$_2$Se$_3$ fixed along $-Z$ direction. The lattice mismatch is no more than $4\%$ for both CrISe and In$_2$Se$_3$ layers with the optimized lattice constants of 3.948 \AA{} for CrISe/In$_2$Se$_3$ heterostructure.
Figures \ref{fig1}(a) and \ref{fig1}(b) show the top and side views of CrISe/In$_2$Se$_3$ heterostructure in the most stable stacking mode (See Supplemental Material for more details~\cite{SupplementalMaterials}). The magnetic atoms Cr form a triangular lattice with the calculated magnetic moment about 3 $\mu_B$ per Cr. Then, the distance between the atomic layers of I and Se$_2$ is defined to be the interlayer distance $L$ as indicated in Fig. \ref{fig1}(b). The weak vdW interaction between CrISe and In$_2$Se$_3$ layers provides convenience for exerting strain vertically, which can be achieved in first-principles calculations by changing the interlayer distance $L$~\cite{RN5281,RN5314}. According to the force acting on atoms I and Se$_2$, we estimate that the necessary pressure for the manipulation of skyrmions and bimerons is in the range of $-$1.1 GPa to 2.7 GPa.

\begin{figure}[!htbp]
	\begin{center}
		\centering
		\includegraphics[width=8.5cm]{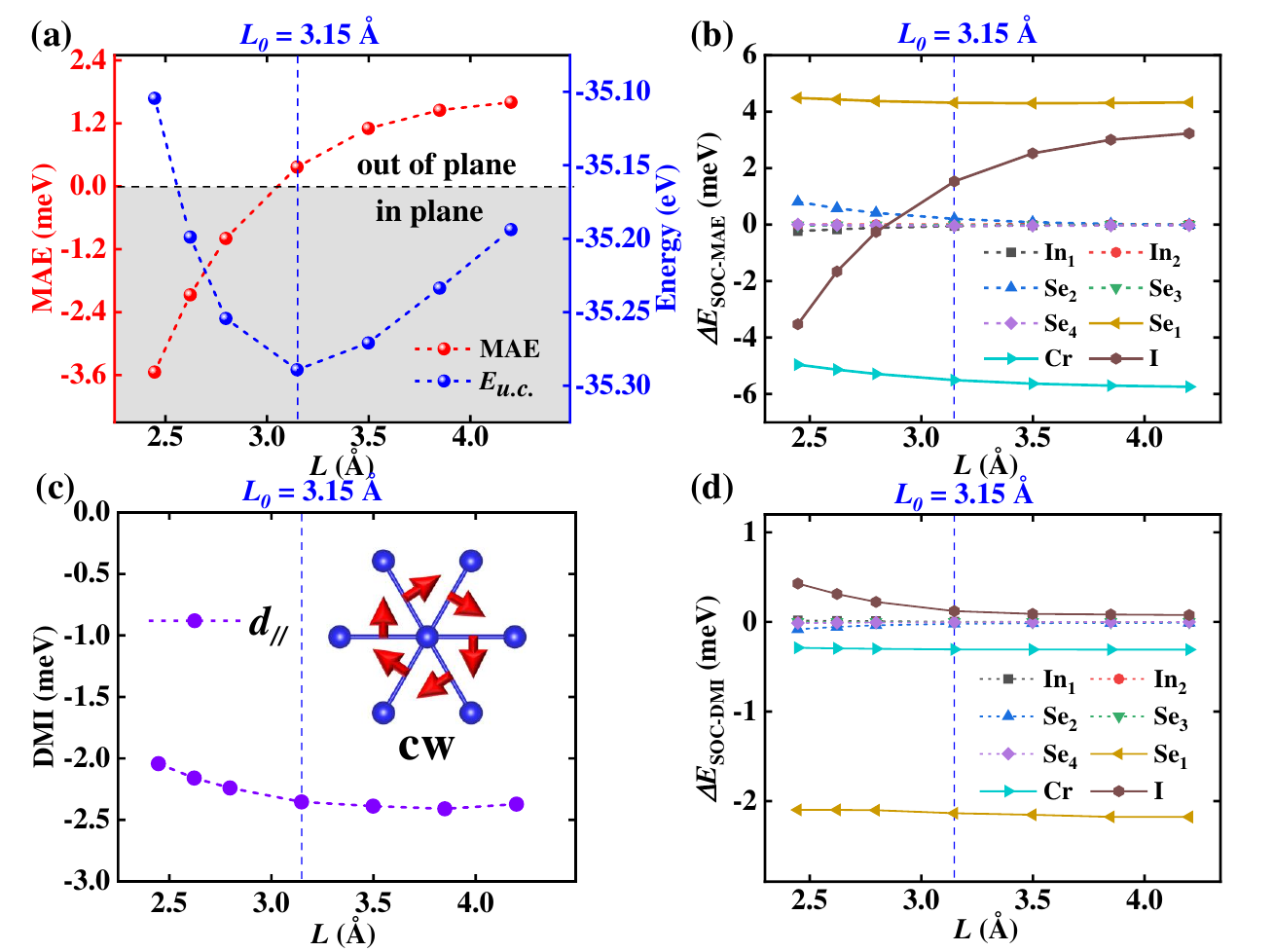}\
		\caption{\label{fig2}
			(a) MAE and the energy of the unit cell ($E_{u.c.}$), (b) atom-resolved SOC energy associated with MAE ($\Delta E_{\rm {SOC\mbox{-}MAE}}$), (c) DMI and (d) atom-resolved SOC energy associated with DMI ($\Delta E_{\rm {SOC\mbox{-}DMI}}$) as functions of the interlayer distance $L$. The optimal interlayer distance $L_0 = 3.15$ \AA{} is indicated by blue dotted lines in (a)-(d). The inset of (c) demonstrates the in-plane components of DMI between the nearest-neighboring Cr atoms.
}
\end{center}
\end{figure}

We first focus on the magnetic anisotropy energy (MAE) of CrISe/In$_2$Se$_3$ heterostructure. According to the Mermin-Wagner theorem~\cite{RN5149}, long range magnetic order is absent in 2D isotropic Heisenberg system with short-range interactions. However, this restriction can be removed by the spin exchange interactions~\cite{RN5459} and MAE~\cite{RN12,RN3000,RN2} in finite-size 2D van der Waals magnets. Here, MAE is defined as the energy difference between the in-plane ($E_x$) and the out-of-plane ($E_z$) ferromagnetic states: MAE = $E_x - E_z$. Thus, a positive (negative) MAE corresponds to the easy-axis (easy-plane) anisotropy. As shown in Fig. \ref{fig2}(a), the optimal interlayer distance of CrISe/In$_2$Se$_3$ is $L_0 = 3.15$ \AA{} at which the energy of the unit cell ($E_{u.c.}$) is lowest. Meanwhile, the MAE of intrinsic CrISe/In$_2$Se$_3$ is 0.4 meV per unit cell with the magnetic easy-axis being out-of-plane. The MAE can be enhanced up to 1.6 meV with increased $L$. Interestingly, as $L$ is reduced, the easy-axis anisotropy switches to easy-plane and the MAE can be further tuned to $-$3.5 meV. This means that the magnetization switching, which is an important operation in spintronic devices especially
for the writing operation, can be achieved by tuning perpendicular strain. Besides, it is noteworthy that the switchable magnetic anisotropy between easy-axis and easy-plane provides the possibility not only for the existence of both skyrmions and bimerons in CrISe/In$_2$Se$_3$ but also for the controllable conversion between them. Figure \ref{fig2}(b) shows the atom-resolved SOC energy associated with MAE ($\Delta E_{\rm {SOC\mbox{-}MAE}}$), from which we can see that the $L$-dependent MAE and the switch of magnetic anisotropy are dominated by I atoms. The Cr and Se$_1$ atoms in CrISe layer also have significant contributions to MAE. However, they are far away from the ferroelectric In$_2$Se$_3$ layer, and thus their contributions to MAE are hardly affected by $L$. The mechanism of the drastically-changed MAE contribution from I atoms can be further elucidated microscopically by the $L$-dependent hybridization of the I-3$p$ orbitals according to the second-order perturbation theory~\cite{RN1969,PhysRevB.95.174424} (See Supplemental Material for more details~\cite{SupplementalMaterials}).

The stability of topological spin textures, such as skyrmions and bimerons is closely related to the Dzyaloshinskii-Moriya interaction (DMI), which stems from SOC and the broken spatial inversion symmetry. To realize nanoscale skyrmions and bimerons, great effort has been made in finding materials with large DMI both experimentally~\cite{RN32,RN5309,RN2982} and theoretically~\cite{RN33,RN3742,RN30,RN66,RN5082}. According to Moriya's rule~\cite{RN131},  since a mirror plane perpendicular to the bond between adjacent Cr atoms passes through the middle of this bond, the DMI between nearest-neighboring Cr is perpendicular to their bond. The DMI vector can be expressed as $\vec{d}_{ij} = d_{//}(\vec{u}_{ij} \times \vec{z}) + d_z\vec{z}$. Here, $\vec{u}_{ij}$ and $\vec{z}$ are the unit vectors from site $i$ to site $j$ and along the $z$ direction, respectively. The influence of $d_z$ is negligible due to its vanishing contribution in average~\cite{RN3705,RN30}. Thus we focus on the $d_{//}$ in the following discussion. As seen in Fig. \ref{fig2}(c), a significant $d_{//}$ of $-$2.35 meV is found in intrinsic CrISe/In$_2$Se$_3$, and its negative value means the clockwise chirality of $d_{//}$ direction denoted by red arrows in the inset. Besides, the $d_{//}$ is almost unaffected by the change of $L$ with the minimum value being $-$2.04 meV, which is stronger than those of the VOI$_2$ (1.76 meV)~\cite{RN3166} and Cr$_2$I$_3$Cl$_3$ (0.38 meV)~\cite{RN5082}. Then, to clarify the microscopic origin of DMI, the $d_{//}$ associated SOC energy $\Delta E_{\rm {SOC\mbox{-}DMI}}$ [the energy difference between clockwise and anticlockwise spin configurations as shown in Fig. S5(a) and S5(b) of the Supplemental Material~\cite{SupplementalMaterials}] is calculated in Fig. \ref{fig2}(d). The dominated contribution to $d_{//}$ mainly stems from the strong SOC of the heavy chalcogen atom Se$_1$. This is similar to the DMI in MnXTe~\cite{RN30} and CrXTe~\cite{RN31} monolayers which can be understood by the Fert-Levy mechanism~\cite{PhysRevLett.44.1538,RN33}. The slightly reduced $d_{//}$ with the decreasing $L$ is mainly caused by the increased $\Delta E_{\rm {SOC\mbox{-}DMI}}$ of I with the opposite sign to that of Se$_1$.

\begin{figure}[!htbp]
	\begin{center}
		\includegraphics[width=8.5cm]{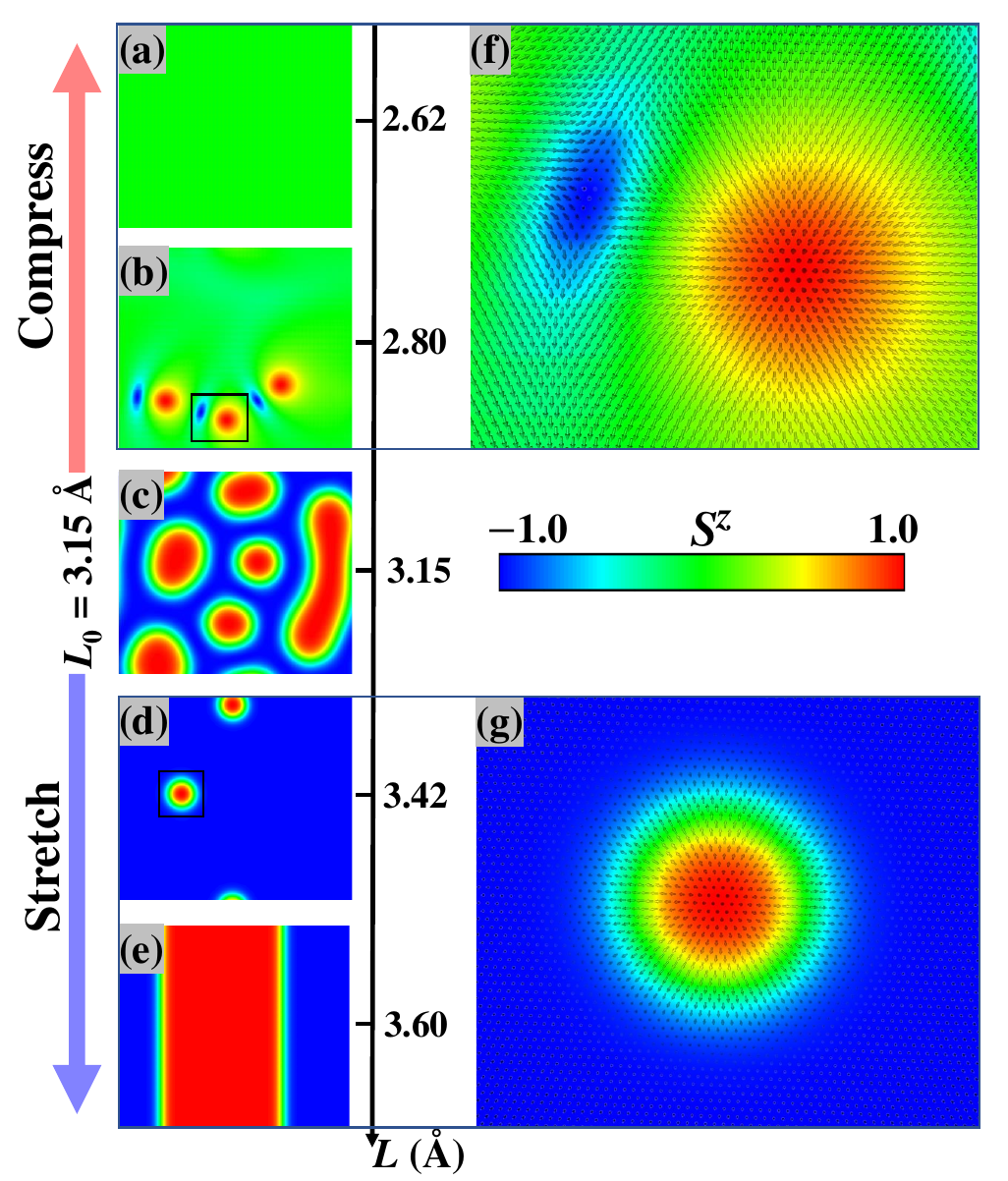}\
		\caption{\label{fig3}
			(a)-(e) Magnetic-field-free spin textures of CrISe/In$_2$Se$_3$ heterostructure with different interlayer distances $L$. (f) and (g) enlarge the framed bimeron and skyrmion in (b) and (d).
		}
	\end{center}
\end{figure}

To explore the topological spin textures in CrISe/In$_2$Se$_3$, atomistic simulations are performed with the magnetic parameters from first-principles calculations. As shown in Fig. \ref{fig3}(c), spontaneous skyrmions with diameters about 14 nm appear in intrinsic CrISe/In$_2$Se$_3$ without external magnetic field. Then, the diameter of skyrmion can be tuned to sub 10 nm by perpendicular stretch as indicated in Fig. \ref{fig3}(d). Figure \ref{fig3}(g) is the enlarged image of a N$\acute{e}$el type skyrmion with a small diameter of 8.0 nm. Notice that skyrmions in such a small size have been widely searched both theoretically and experimentally due to the promising application in low-energy-consumption and high-density memory devices~\cite{RN137,RN121,RN3528,RN3177,RN3042}. Then, as shown in Fig. \ref{fig3}(e), with the further increase of $L$, the skyrmions are annihilated to uniform ferromagnetic domains separated by N$\acute{e}$el type domain walls.
More interestingly, as shown in Fig. \ref{fig3}(b), the magnetic-field-free conversion from skyrmions to bimerons is observed under perpendicular compression due to the switch of magnetic anisotropy from easy-axis to easy-plane, accompanied by the magnetization switching from out-of-plane to in-plane. Figure \ref{fig3}(f) shows the zero-magnetic-field bimeron consisting of a meron ($Q = 0.5$) and an antimeron ($Q = 0.5$) with mutually reversed out-of-plane magnetizations. Here, $Q$ is the topological charge defined as~\cite{RN3688}: $Q = \frac{1}{4\pi}\int \vec{S}\cdot (\partial_x \vec{S} \times \partial_y \vec{S})dxdy$. Besides, similar to the case of skyrmions, size modulation of bimerons can be realized by further compress. Notice that, without rotational symmetry, bimerons are highly promising for racetrack applications. But so far, potential bimeron systems are much rarer than those of the skyrmions due to the exacting requirement of DMI~\cite{RN2009,RN3648} and easy-axis magnetic anisotropy of most 2D vdW magnets~\cite{RN3764,RN11,RN61}. Additionally, the bimerons in CrISe/In$_2$Se$_3$ can not only be produced, but also be annihilated to an in-plane ferromagnetic state by perpendicular compression as shown in Fig. \ref{fig3}(a). Thus, the minimum bits `1' and `0' for information storage can be achieved in CrISe/In$_2$Se$_3$ via the generation and annihilation of bimerons or skyrmions simply by a perpendicular strain without any external field. Notice that these metastable spin textures are not due to the frustration factor |$J_3$/$J_1$| which is very small ($\sim$0.05), but due to the DMI and topological robustness of individual skyrmion or bimeron. We further calculate the energy difference between the spin textures in Fig. 3(b) [3(d)] and in-plane (out-of-plane) ferromagnetic state, which is 44 meV/spin (30 meV/spin).

\begin{figure}[!htbp]
	\begin{center}
	\includegraphics[width=8.5cm]{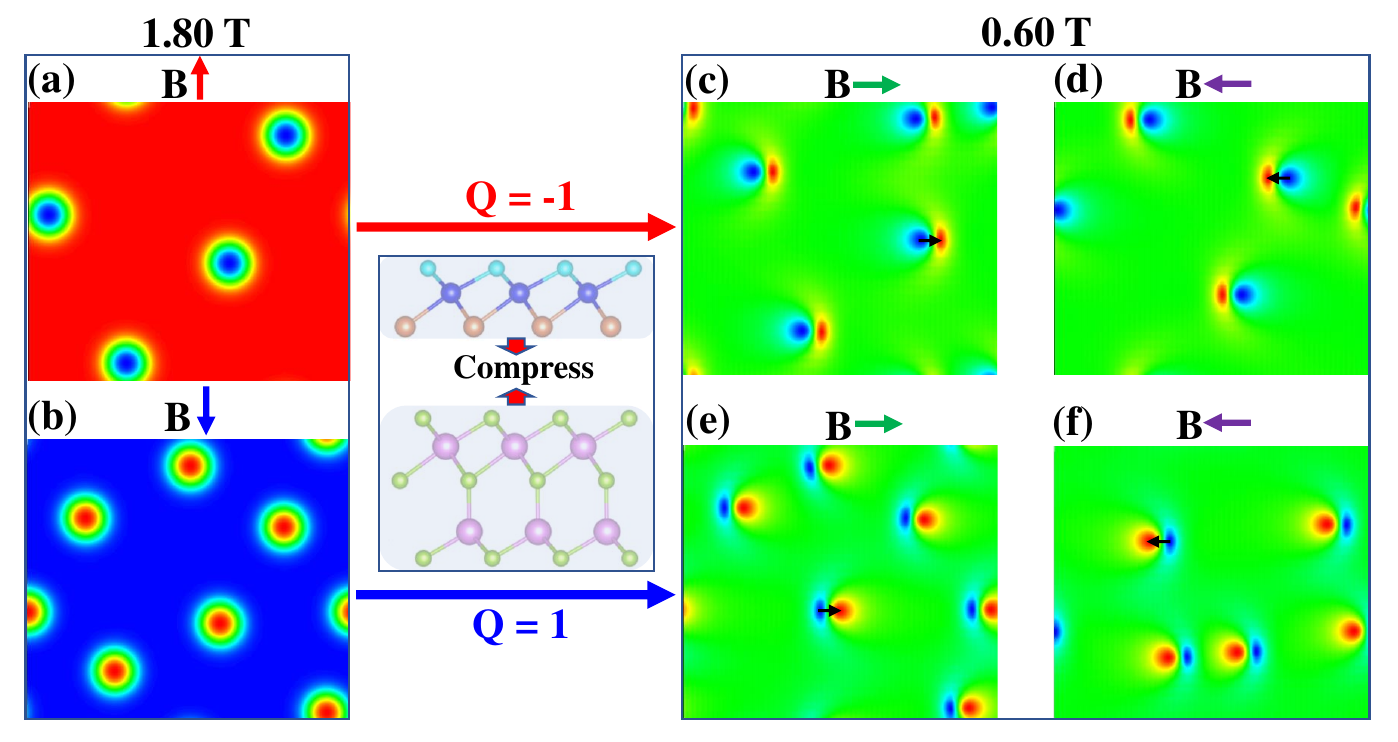}\
	\caption{\label{fig4}
	The compression induced conversion from skyrmions to bimerons with controllable topological charge and polarity. (a), (c) and (d) $Q = -1$, (b), (e) and (f) $Q = 1$. The black arrows in (c)-(f) indicate the polarity of bimerons ($\vec{p}$), which is a unit vector from the center of spin down to the center of spin up. The color map is the same as that of Fig. \ref{fig3}.
	}
	\end{center}
\end{figure}

The topological charge $Q$ is a fingerprint for topological property that determines the topologically protected stability, and furthermore the topology-related other characteristics. For example, the sign and magnitude of skyrmion Hall angle is directly related to $Q$~\cite{RN2982,RN5310}. Thus, the direction of lateral displacement in which skyrmion deviates from the driving force can be controlled by restricting the sign of $Q$. Therefore, the control of $Q$ plays an important role in the manipulation of topological textures, which can be realized in  CrISe/In$_2$Se$_3$. As shown in Fig. \ref{fig4}(a) and \ref{fig4}(b), $Q$ can be restricted as $+$1 ($-$1) with a magnetic field of 1.8 T along $+$Z ($-$Z) direction. Then, by applying perpendicular compression and a small in-plane magnetic field about 0.6 T, skyrmions transform into bimerons with $Q$ unchanged. Besides, we define the polarity of bimeron by a unit vector pointing from the spin down center to the spin up one as shown by the black arrows in Fig. \ref{fig4}(c)-\ref{fig4}(f). It can be seen that the polarity vector of bimeron is always in the same direction of the uniform background magnetization, which can be controlled by the direction of external magnetic field. Notice that the speed of bimerons could be regulated by tuning the angle between external drive and the background magnetization~\cite{RN5310}, which means the possibility to manipulate the polarity related dynamic characteristics of bimeron via external magnetic field in similar magnetic systems. Therefore, the bimeron with certain $Q$ and polarity can be obtained in CrISe/In$_2$Se$_3$ by perpendicular strain and magnetic field, which provides a superior platform for dynamic manipulation.

\begin{figure}[!htbp]
	\begin{center}
		\includegraphics[width=8.5cm]{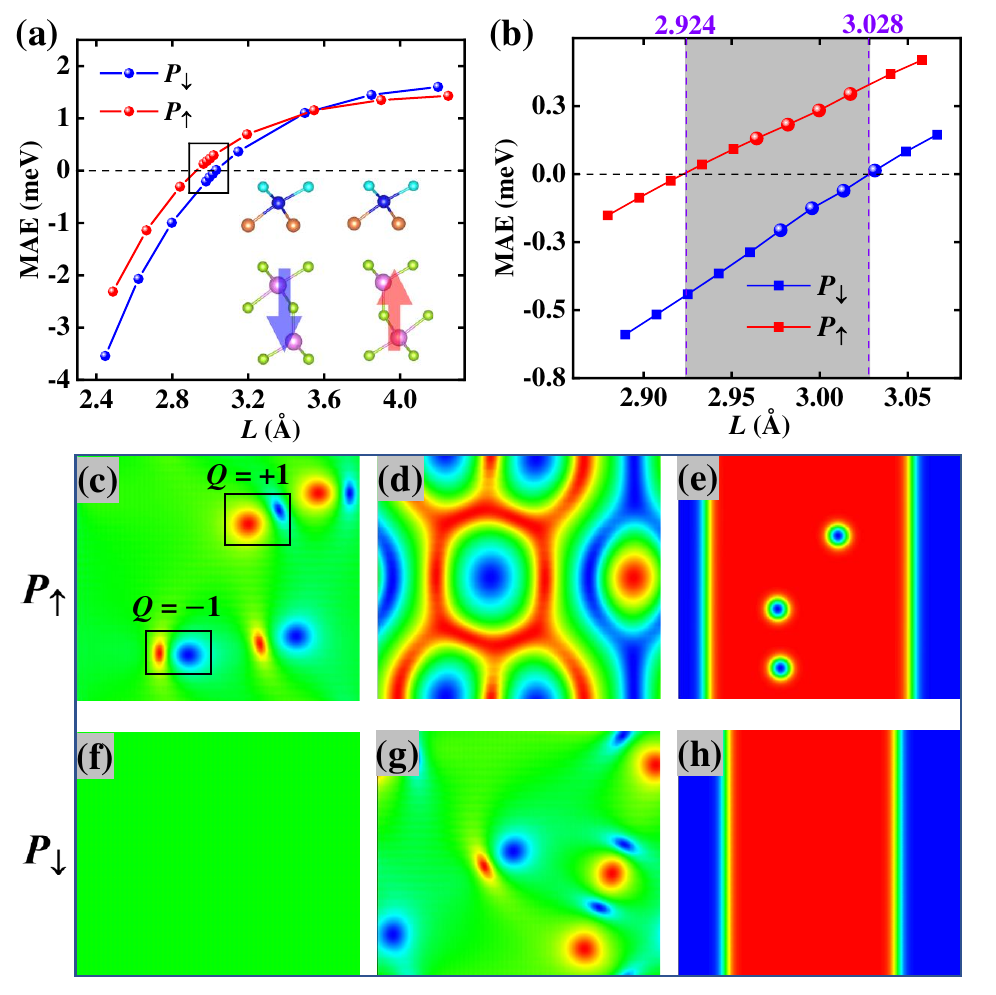}\
		\caption{\label{fig5}
		Regulation of MAE and magnetic topological textures by polarization. (a) MAE of CrISe/In$_2$Se$_3$ heterostructure with opposite polarization of In$_2$Se$_3$ layer as a function of interlayer distance $L$. (b) The enlarged range of $L$ where the magnetic anisotropy can be tuned between easy-axis and easy-plane by polarization. The round and square points in (b) correspond to the results by DFT and interpolation, respectively. (c) - (h) Magnetic-field-free spin textures of CrISe/In$_2$Se$_3$ with the polarization of In$_2$Se$_3$ along $+Z$ [(c) - (e)] and $-Z$ [(f) - (h)]. $L$ are 2.63 [(c), (f)], 2.77 [(d), (g)], and 3.88 [(e), (h)] \AA{} respectively. The color map of (c) - (h) is the same as that of Fig. \ref{fig3}.
		}
	\end{center}
\end{figure}
In recent years, manipulation of topological magnetic textures via electric field begins to attract considerable attention due to the nearly dissipation-free advantage~\cite{RN3166,RN5244}. Here, taking advantage of the reversible and nonvolatile ferroelectric polarization in In$_2$Se$_3$ layer, the magnetic properties of CrISe layer can be manipulated by electric field in this heterostructure. The calculation indicates that CrISe/In$_2$Se$_3$ with $P$ of In$_2$Se$_3$ fixed along $+Z$ direction ($P_{\uparrow }$) stabilizes in the same stacking mode as CrISe/In$_2$Se$_3$($P_{\downarrow }$). As shown in Fig. \ref{fig5}(a), as functions of interlayer distance $L$, the MAE of CrISe/In$_2$Se$_3$($P_{\uparrow }$) shows similar trend as that of CrISe/In$_2$Se$_3$($P_{\downarrow }$), and the switch of magnetic anisotropy from easy-axis to easy-plane could be achieved in both cases. Interestingly, the change of MAE with $L$ is not synchronous for CrISe/In$_2$Se$_3$($P_{\uparrow }$) and CrISe/In$_2$Se$_3$($P_{\downarrow }$), which enables us to tune the magnetic anisotropy and the topological spin textures via ferroelectric polarization. Figure. \ref{fig5}(b) displays the details around MAE = 0 meV, from which we can see that in the shadow region (2.924 \AA{} $< L <$ 3.028 \AA{}), the magnetic anisotropy can be switched between easy-axis and easy-plane by tuning the polarization of In$_2$Se$_3$. Besides, as shown in Fig. \ref{fig5}(c) ($L = 2.63$ \AA{}), bimerons with opposite $Q$s coexist in CrISe/In$_2$Se$_3$($P_{\uparrow }$) with the easy-plane magnetic anisotropy of $-$1.3 meV. Nevertheless, when the polarization of In$_2$Se$_3$ is reversed to the $P_{\downarrow }$ state, the bimerons disappear accompanied by a vanishing $Q$ [Fig. \ref{fig5}(f)]. When $L$ increases to 2.77 \AA{}, skyrmions with opposite $Q$s appear in CrISe/In$_2$Se$_3$($P_{\uparrow }$). Due to the weak easy-plane anisotropy ($-$0.6 meV), the in-plane magnetization part (green area) of skyrmions is wider than the skyrmions in easy-axis magnets. Similar spin textures are also demonstrated in MnBi$_2$Se$_2$Te$_2$/In$_2$Se$_3$~\cite{RN5244}. Then, by switching the polarization of In$_2$Se$_3$ to $P_{\downarrow }$, the easy-plane anisotropy is enhanced to $-$1.1 meV, resulting the conversion from skyrmions to bimerons [Fig. \ref{fig5}(g)].
Moreover, with the $L$ further increasing to 3.88 \AA{}, the magnetic anisotropy transforms to easy-axis and sub-5 nm skyrmions appear in CrISe/In$_2$Se$_3$($P_{\uparrow }$). While the skyrmions are annihilated in CrISe/In$_2$Se$_3$($P_{\downarrow }$) due to the large MAE of 1.5 meV. Thus, we demonstrate that the creation and annihilation of both skyrmions and bimerons, as well as the conversion between them can be realized in CrISe/In$_2$Se$_3$ by applying electric field to reverse the ferroelectric polarization of In$_2$Se$_3$. Such magnetic-field-free multi-functional control of magnetic topological textures by polarization is highly desired in nonvolatile data storage with low-energy consumption.

\textit{Conclusion}. 
In conclusion, by first-principles calculations and atomistic simulations, we predict vdW magnetoelectric heterostructure CrISe/In$_2$Se$_3$ to be an ideal platform for magnetization switching and the multi-functional manipulation of magnetic topological textures in energy-efficient ways. By tunning perpendicular strain, the MAE can be tunned in a wide windows from $-$3.5 meV to 1.6 meV, while the magnetic anisotropy can be switched between in-plane and out-of-plane. Thereby, the generation and annihilation of skyrmions and bimerons, the conversion between them as well as the size modulation could all be manipulated in this magnetic-field-free way. Moreover, the multi-functional manipulation of skyrmions and bimerons can also be achieved by applying electric field to tune the ferroelectric polarization of In$_2$Se$_3$. Our study predicts CrISe/In$_2$Se$_3$ as a promising candidate for spintronic applications, guiding the way for low-dissipation manipulation of magnetic topological textures via strain and polarization.

\begin{acknowledgments}
This work is supported by the Natural Science Foundation of Jiangsu Province (BK20221451), the National Natural Science Foundation of China (11834002, 12274070). Most calculations were done on the Big Data Computing Center of Southeast University.
\end{acknowledgments}

\bibliography{ref}
\bibliographystyle{apsrev4-2}
\end{document}